\documentclass[aps,prl,preprint,groupedaddress]{revtex4}

\input{epsf}

\begin{document}

\title{\textbf{Evidence of mobile carriers with Charge Ordering gap in Epitaxial Pr$_{0.625}$Ca$_{0.375}$MnO$_{3}$ Thin Films}}


\author{Udai Raj Singh, S. Chaudhuri, Anirban Dutta, R. C. Budhani, and Anjan K. Gupta}
\affiliation{Department of Physics, Indian Institute of Technology Kanpur, Kanpur 208016, India.}
\date{\today}

\begin{abstract}
Epitaxial thin films of charge-ordered Pr$_{0.625}$Ca$_{0.375}$MnO$_{3}$ have been studied using variable temperature Scanning tunneling microscopy and spectroscopy (STM/STS). The as grown films were found to be granular while the annealed films show atomic terraces at all temperatures and are found to be electronically homogeneous in 78-300K temperature range. At high temperatures (T$>$T$_{CO}\approx$ 230 K) the local tunnel spectra of the annealed films show a depression in the density of states (DOS) near Fermi energy implying a pseudogap with a significant DOS at E$_F$.  The gap feature becomes more robust with cooling with a sharp jump in DOS at E$_F$ at T$_{CO}$ and with a gap value of $\sim$0.3 eV at 78K. At low temperatures we find a small but finite DOS at E$_F$ indicative of some delocalized carriers in the CO phase together with an energy gap. This is consistent with bulk transport, which shows weakening of the activation gap with cooling below 200K, and indicates the presence of two types of carriers at low temperatures.

\end{abstract}


\maketitle
The narrow bandwidth charge-ordered (CO) manganites have been of significant research interest for their intriguing physics \cite{salamon} and application potential in memory applications as the insulating CO phase can be melted into metallic phase by various perturbations such as magnetic field, electric field and electromagnetic radiation \cite{Kiryukhin,Okimoto,Padhan}. The manganites, in general, are best understood by Zener double-exchange mechanism \cite{Zener} with modifications arising from electron-lattice coupling due to Jahn-Teller (JT) interaction \cite{Millis,hwang-batlogg}. The relative strength of the localizing JT interaction as compared to the hopping energy is an important parameter determining the electron mobility and bandwidth \cite{Millis}. The ferromagnetic ordering promotes hopping of carriers due to double exchange while an antiferromagnetic ordering is expected to suppress it \cite{Millis}. In narrow bandwidth (BW) manganites, like Pr$_{1-x}$Ca$_{x}$ MnO$_{3}$ (PCMO), the JT interactions dominate over hopping leading to the carrier trapping into localized JT polarons. These polarons hop with an activation energy as seen in the bulk transport at high temperatures \cite{polaron-hopping,salamon}.  Due to these strong JT interactions, PCMO does not show the insulator-metal transition under normal conditions. In particular, for the doping range 0.30$\leq$x$\leq$0.65, PCMO shows a CO transition in 170-240K temperature range with cooling \cite{Tokura}.

In recent studies of PCMO manganites, the CO transition, structural transition and long range antiferromagnetic ordering have been investigated extensively by various experiments such as X-ray scattering, neutron scattering, optical conductivity and angle resolved photoemission spectroscopy (ARPES). The resonant X-ray scattering and neutron diffraction experiments reveal simultaneous structural and CO transitions \cite{Kajimoto,Zimmermann,Grenier,Cox,Jirak}. A long range CE-type antiferromagnetic ordering has also been found in bulk PCMO below T$_{N}$ =170 K which is slightly below its T$_{CO}$ \cite{Kajimoto,Grenier,Cox}. Thus the ground state of PCMO is CO with CE-type antiferromagnetic ordering \cite{Kajimoto,Grenier}. In photoemission and optical measurements of PCMO, the suppression of spectral weight around T$_{CO}$  shows a clear evidence of the CO transition \cite{Ebata,Okimoto}. The STM measurements of three dimensional CO manganites like Nd$_{0.50}$Sr$_{0.50}$MnO$_{3}$ and Bi$_{0.24}$$Ca_{0.76}$Mn$_{3}$ show a fully opened hard gap on the surface \cite{renner-cheong,Biswas,Kar} in the CO phase. Renner et. al. have reported charge-ordering, with a hard gap, and charge disordered state with atomic resolution in layered Bi$_{0.24}$$Ca_{0.76}$MnO$_{3}$ single crystals \cite{renner-cheong}. In a bilayer manganite a hard gap is seen throughout a broad temperature range across insulator-metal transition \cite{renner-bilayer}.

In this paper we present the temperature dependent scanning tunneling microscopy and spectroscopy (STM/STS) of as grown and annealed epitaxial Pr$_{0.625}$Ca$_{0.375}$MnO$_{3}$ thin films from 295 to 78K. As opposed to the as grown films, which are granular, the annealed films show atomically flat terraces with homogeneous tunneling density of states (DOS) at all temperatures with a pseudogap evolving into a more robust gap with cooling below T$_{CO}$. More interestingly, the bulk resistivity of the films shows a reduction in activation gap with cooling below 200 K. We discuss these results in terms of the existence of states at E$_F$ together with a CO gap.

\begin{figure}
\epsfxsize = 3.2 in \epsfbox{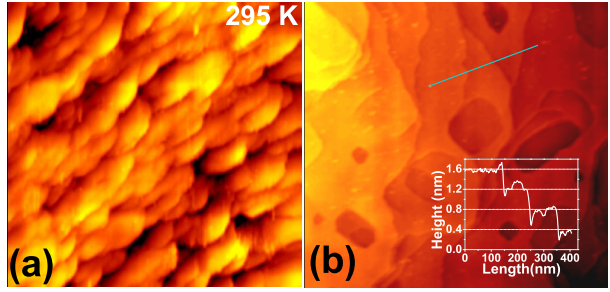}
\caption{\label{fig:PCMO-topo} (a)Topographic image of as-grown Pr$_{0.625}$Ca$_{0.375}$MnO$_{3}$ films with scan size 0.607$\mu$m$\times$0.607$\mu$m at 295 K. Tunneling parameters are 1.0 V bias voltage and 0.2 nA tunnel current. (b) Topographic image of an annealed PCMO film  taken at 295 K (bias = 1.0 V, I = 0.12 nA, and image size = 1.24$\mu$m$\times$1.24$\mu$m).}
\end{figure}

Strain free epitaxial Pr$_{0.625}$Ca$_{0.375}$MnO$_{3}$ thin films were grown on (110) surface of NdGaO$_3$ substrates using laser ablation technique. These films were transferred to the STM cryostat with a very short exposure (${<}$30 min) to the air. The topographic STM images of as grown films show a granular structure (see Fig.\ref{fig:PCMO-topo}a) with an rms surface roughness of $\sim$10 nm. The STS conductance maps of the as grown films show marked electronic inhomogeneity that correlates well with the granular structure with the grain boundaries showing more insulating behavior than the grains. The spectra also show an enhancement in gap behavior with cooling; however, the changes in spectra are less sharp with cooling as compared to the annealed films, which are discussed further. The details of STM/S measurements have been discussed in an earlier publication \cite{Udai}. For conductance imaging, an ac modulation voltage of 50 mV and frequency 2.571kHz was applied over the DC bias and corresponding modulation in current was detected using a lock-in amplifier. All the spectra reported here were acquired at a junction resistance of 10 G$\Omega$ (1.0 V bias and 0.1 nA tunnel current). The zero bias conductance was found from the I-V spectra directly to avoid artifacts arising from the effects related to tip-sample capacitance in the ac-modulation technique.

\begin{figure}
\epsfxsize = 3.4 in \epsfbox{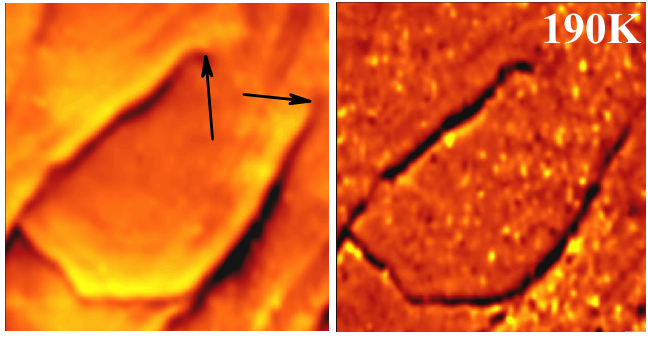}
\caption{\label{fig:PCMO-cond} Topographic and conductance images (size 0.372$\mu$m$\times$0.401$\mu$m) of PCMO film at 190 K taken at 1.0 V dc bias with 0.10 nA tunnel current and an ac modulation voltage of 50mV.}
\end{figure}

Recent X-ray absorption spectroscopy shows that the annealing of the manganite films in air helps to rule out the formation of Mn$^{2+}$ ions and oxygen deficiency with no effect on transition temperatures \cite{Valencia}. Annealing also makes the surface atomically smooth \cite{udai-apl}. Thus we annealed the as grown PCMO films at 800$^{0}$C for eight hours in air to obtain a homogeneous and smooth surface and to ensure that we are probing the intrinsic surface. We present a topographic image of the annealed Pr$_{0.625}$Ca$_{0.375}$MnO$_{3}$ film taken at 295 K in Fig.\ref{fig:PCMO-topo}b. The line scan of 295 K image in the inset of Fig.\ref{fig:PCMO-topo}b shows flat terraces separated by atomic height (0.40 nm $\pm$ 0.05 nm) steps. Such homogeneous film surface is more appropriate for probing intrinsic electronic inhomogeneities and variation in local DOS. We see similar terraces in the studied temperature range (295-78 K). The terraces in annealed films are formed due to diffusion of constituents at very high temperatures as opposed to the step-flow growth mode in films deposited at high temperature with very slow deposition rate. The latter type films show a uniform step terrace morphology \cite{Udai}. Sometimes, we also see steps with multiple unit-cell height and some screw dislocations (marked in Fig.\ref{fig:PCMO-cond}a).

\begin{figure}
\epsfxsize = 3.0 in \epsfbox{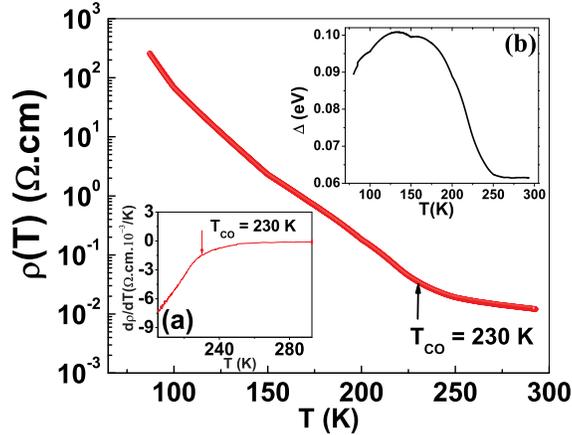}
\caption{\label{fig:PCMO-res} The resistivity measurements of annealed Pr$_{0.625}$Ca$_{0.375}$MnO$_{3}$ films. The CO transition, at $T_{CO}$=230 K, is clearly observable in the derivative plot of resistivity shown in the inset (a). Inset (b) shows the temperature dependent variation of activation gap ($\Delta(T)$) value (see text for details). $\Delta(T)$ also shows a sharp jump near $T_{CO}$.}
\end{figure}

The two probe resistivity measurement on these annealed films was carried out using a closed cycle helium refrigerator after performing the STM measurements. The $\rho$(T) data plotted in Fig.\ref{fig:PCMO-res} show a sudden rise in resistivity around 230 K which we attribute to the onset of CO. This is more clear in the derivative plot shown in inset {\bf a} of Fig.\ref{fig:PCMO-res}. These Pr$_{0.625}$Ca$_{0.375}$MnO$_{3}$ films on NdGaO$_{3}$ substrates are almost strain free with T$_{CO}$ = 230 K, nearly equal to the bulk T$_{CO}$ of 235 K in Pr$_{0.600}$Ca$_{0.400}$MnO$_{3}$ single crystals \cite{Okimoto}. We also note that the CO temperature is not affected significantly by annealing. The resistivity behavior is that of a gapped insulator, i.e. $\rho(T)=\rho(T_0)exp\left\{\frac{\Delta(T)}{k{_B}T}-\frac{\Delta(T_0)}{k{_B}T_0}\right\}$. Here, $\Delta$ is the activation gap while $\rho(T_0)$ is resistivity value at a given temperature $T_0$. For T$>$250K we can fit the resistivity to this form with a constant activation gap ($\Delta$) of 0.062eV. For lower temperatures, we extract the temperature dependent activation gap value using relation $\Delta(T)=k{_B}T\left[\frac{\Delta({T_0})}{k{_B}T{_0}}+ln\left\{\frac{\rho(T)}{\rho(T{_0})}\right\}\right]$, as derived from the activated resistivity. We have taken $T_0$=292K with $\Delta(T_0)$=0.062eV. This $\Delta(T)$ is plotted in the inset {\bf b} of Fig.\ref{fig:PCMO-res} showing a rise in the gap value from 0.062eV above $T_{CO}$ to 0.1eV with cooling, which is $\sim$5$k_BT_{CO}$ \cite{res-note}. Below 125K the gap starts to decrease, which may be interpreted either as a reduction of activation gap or, perhaps, another conduction mechanism beyond the activated behavior if the gap does not decrease. As discussed later from the tunneling spectra, we do not find a reduction in gap with cooling and in fact some states remain at the Fermi energy at low temperatures together with a gap; this we believe is responsible for the low temperature resistivity behavior.

Fig.\ref{fig:PCMO-cond} shows the simultaneous topographic and conductance images of an annealed film at 190K. The conductance shows little variation in the local conductance on each of the terrace except for some defects which we believe arise from some chemical inhomogeneity in these non-stoichiometric films. There is a jump in conductance at the steps resulting from the feedback instability at the steps due to small feedback bandwidth. We see similar homogeneous conductance images at all the studied temperatures and thus we conclude that the film surface is electronically homogeneous.

\begin{figure}
\epsfxsize =3.2 in \epsfbox{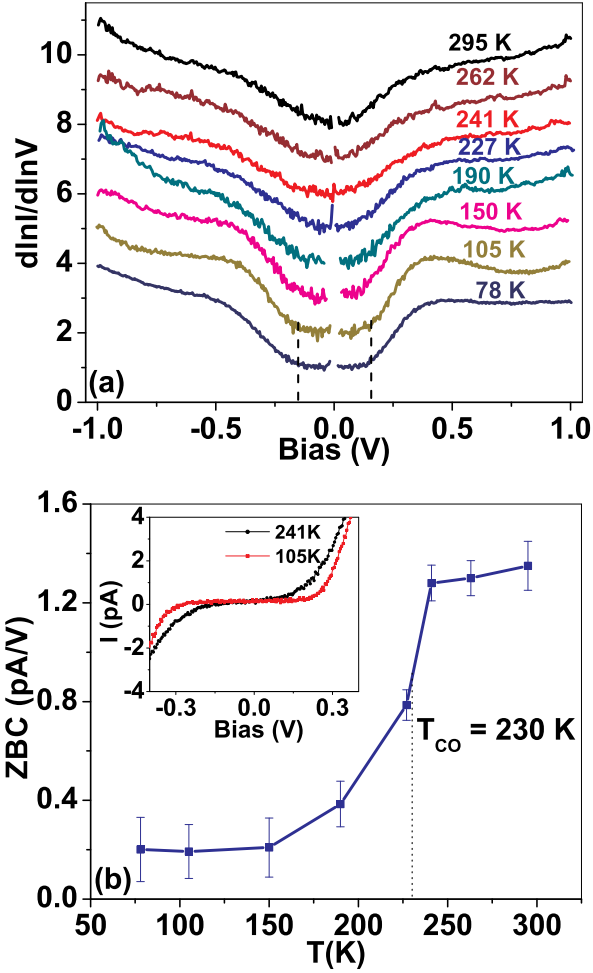}
\caption{\label{fig:Plot-PCMO} (a) Temperature variation of tunnel spectra of annealed films taken with tunnel parameters of 1.0 V bias and 0.1 nA tunnel current. The spectral evolution from pseudogap to a hard gap indicates the CO transition around 230K. The dashed lines show an interval of 0.3eV indicating the scale of the energy gap. (b) The temperature variation of ZBC from 295K to 78K as found from the zero bias slope of the I-V spectra. Two I-V spectra are also shown in the inset.}
\end{figure}

The temperature dependant tunneling spectra are shown in Fig.\ref{fig:Plot-PCMO}a. Here, we have plotted dlnI/dlnV vs V to normalize away the voltage dependence of the tunneling matrix element \cite{normalize-ref}.  Each plotted spectrum at a particular temperature is an average of about hundred spectra taken at different points of the sample surface to average over small variations. At 295K, there is a V-shaped dip in the bias range of $\pm$0.1 V with a weakly rising background in the spectra reflecting a pseudogap with non-zero DOS at E$_F$. Similar pseudogap has also been observed in PCMO by photoemission and optical studies \cite{Ebata,Liu}. This pseudogap becomes more pronounced with cooling and a more robust gap starts appearing near 230K with cooling, which is qualitatively consistent with the resistivity activation behavior discussed earlier.

By definition, the value of dlnI/dlnV is one at V = 0 so it can not be used to make out the variations in DOS at E$_F$. Therefore, we plot the zero bias conductance (ZBC), i.e. dI/dV at V = 0, as a function of temperature in Fig.\ref{fig:Plot-PCMO}b. These ZBC values were found from the slope at V = 0 of spatially averaged I-V spectra taken at a fixed junction resistance of 10 G$\Omega$ at different temperatures. Two representative I-V spectra are also shown in the inset of Fig.\ref{fig:Plot-PCMO}b. As seen in Fig.\ref{fig:Plot-PCMO}a, the energy scale of the dip does not change much with cooling at T$_{CO}$; however there is a marked change in the behavior of ZBC.

We estimate the magnitude of the energy gap (2$\Delta$) from separation between the two gap edges to be $\sim$0.3 eV below 230 K. This is of the similar magnitude but slightly larger than the maximum resistivity activation gap, $\Delta$= 0.1eV (see Fig.\ref{fig:PCMO-res} inset b). This magnitude is also consistent with the optical conductivity measurements \cite{Okimoto}. With further cooling, a more robust gap appears from 150K downwards with a marked change in the behavior of the spectra (see Fig.\ref{fig:Plot-PCMO}a) with sharper gap edges and nearly zero curvature near V = 0. We can also see a maxima in the normalized conductance just outside the positive bias gap edge for 150K and lower temperatures. The temperature below which a more robust gap appears is close to the antiferromagnetic transition temperature, T$_N$ $\approx$ 170K of bulk PCMO \cite{Kajimoto}. However, we have not confirmed such antiferromagnetic ordering in our thin films.

From Fig.\ref{fig:Plot-PCMO}b we see that ZBC does not go to zero within the error bars indicating some carriers at E$_{F}$ at low temperatures in CO phase. These error bars are found from the spread in ZBC as a function of position on the surface. Although general behavior of ZBC and the gapped spectra agree with the bulk transport and makes us believe that we are probing the intrinsic bulk of the sample, the possibility of surface behaving differently from the bulk can not be ruled out completely, as seen to be the case with some of the manganites \cite{Park-Freeland,Udai}. Now if we compare the temperature evolution of the tunnel spectra and resistivity activation energy (Fig.\ref{fig:PCMO-res}b), we see that the activation energy above 260K is constant but small ($\sim$0.06eV) possibly giving rise to a thermally smeared (pseudo) gap-like feature in the tunnel spectra. We cannot distinguish between actual DOS at E$_F$ from the thermally activated DOS at this temperature. With cooling, the rise in resistivity activation gap is consistent with decrease in ZBC together with a more robust energy gap in tunnel spectra. Below 125 K, the activation gap decreases while the tunneling gap remains same (or even becomes stronger). However, the ZBC (or DOS at E$_F$) never goes to zero within the error bars even at the lowest temperatures. 

The temperature dependence of ZBC cannot be merely due to thermal activation of carriers across the gap as we can not fit an activated behavior of the type $Exp(-\Delta/k_{B}T)$ for ZBC for T $<$ T$_{CO}$ as the gap energy scale of $\sim$0.15 eV is much larger than the thermal energy available at low temperatures. This CO is also known to be very susceptible to perturbations and can be easily melted \cite{Kiryukhin,Okimoto,Padhan} into metallic phase. Further, the ferromagnetic correlations \cite{Cox} seen with neutron studies in PCMO without any perturbations can also delocalize carriers by double exchange mechanism and thus would be consistent with the states inside the gap.

In broad bandwidth bilayer manganites, local (in k-space) pockets of quasi-particles were found, together with an energy gap, and were attributed to coherent polarons \cite{coh-pol}. Our recent work on relatively broad bandwidth manganites LSMO \cite{udai-apl} and LCMO \cite{udai-jpcm} also showed a pseudogap with an increase in DOS at E$_F$ with cooling. These states could arise either from extended (or coherent) polarons \cite{coh-pol} or from dynamic phase fluctuations \cite{dagotto}. In dynamic phase separation, one can have conducting regions and the CO insulating regions fluctuating on a very short time scale giving a time averaged DOS having signatures of both phases on STM measurement time scales.

In conclusion, we find that the CO transition, observed by bulk transport in epitaxial Pr$_{0.625}$Ca$_{0.375}$MnO$_{3}$ films at 230 K is marked by the opening of CO gap in tunnel spectra. A depression in DOS above T$_{CO}$ can be interpreted as a polaronic pseudogap, which evolves into a more pronounced CO gap with cooling. From the qualitative agreement of temperature dependence of the activation gap found from bulk transport with the finite DOS at $E_F$ in tunnel spectra we believe that there are some delocalized states that coexist with the gapped charge-ordered states at low temperatures.

URS acknowledges financial support from CSIR. AKG and RCB acknowledge research grants from DST of the Govt. of India and the Indo-French centre for promotion of Advanced Research, respectively. We thank P. K. Rout for doing the resistivity measurements of annealed PCMO films.


\begin{thebibliography}:
\bibitem{salamon} M. B. Salamon and M. Jaime, Rev. Mod. Phys. {\bf 73}, 583 (2001) and references therein.
\bibitem{Kiryukhin} V. Kiryukhin, D. casa, J. P. Hill, B. Keimer, A. Vigliante, Y. Tomoika, and Y. Tokura, Nature {\bf 386}, 813 (1997).
\bibitem{Okimoto} Y. Okimoto, Y. Tomioka, Y. Onose, Y. Otsuka, and Y. Tokura, Phys. Rev. B {\bf 57}, R9377 (1998).
\bibitem{Padhan} P. Padhan, W. Prellier, Ch. Simon, and R. C. Budhani, Phys. Rev. B {\bf 70}, 134403 (2004).
\bibitem{Zener} C. Zener, Phys. Rev. {\bf 82}, 403 (1951).
\bibitem{hwang-batlogg} H. Y. Hwang, S-W. Cheong, P. G. Radaelli, M. Marezio, and B. Batlogg, Phys. Rev. Lett. {\bf 75}, 914 (1995).
\bibitem{Millis}A. J. Millis, B. I. Shraiman, and R. Mueller, Phys. Rev. Lett. {\bf 77}, 175 (1996).
\bibitem{polaron-hopping} M. Jaime, M. B. Salamon, M. Rubinstein, R. E. Treece, J. S. Horwitz and D. B. Chrisey, Phys. Rev. B {\bf 54}, 11914 (1996).
\bibitem{Tokura} Y. Tokura and Y. Tomioka, J. Magn. Magn. Mater. {\bf 200}, 1 (1999).
\bibitem{Jirak} Z. Jir\'{a}k, S. Krupicka, Z. Simsa, M. Dlouh\'{a}, and S. Vratislav, J. Magn. Magn. Mater. {\bf 53}, 153 (1985)
\bibitem{Kajimoto} R. Kajimoto, T. Kakeshita, Y. Oohara, H. Yoshizawa, Y. Tomioka, and Y. Tokura, Phys. Rev. B {\bf 58}, R11837 (1998)
\bibitem{Zimmermann} M. V. Zimmermann, J. P. Hill, D. Gibbs, M. Blume, D. Casa, B. Keimer, Y. Murakami, Y. Tomioka, and Y. Tokura, Phys. Rev. Lett. {\bf 83}, 4872 (1999), M. V. Zimmermann, C. S. Nelson, J. P. Hill, D. Gibbs, M. Blume, D. Casa, B. Keimer, Y. Murakami, C.-C. Kao, C. Venkataraman, T. Gog, Y. Tomioka, and Y. Tokura, Phys. Rev. B {\bf 64}, 195133 (2001)
\bibitem{Grenier} S. Grenier, J. P. Hill, D. Gibbs, K. J. Thomas, M. V. Zimmermann, C. S. Nelson, V. Kiryukhin, Y. Tokura, Y. Tomioka, D. Casa, T. Gog, and C. Venkataraman, Phys. Rev. B {\bf 69}, 134419 (2004).
\bibitem{Cox} D. E. Cox , P. G. Radaelli ,M. Marezio,  and  S-W. Cheong , Phys. Rev. B {\bf 57} 3305 (1998).
\bibitem{Ebata} K. Ebata, M. Hashimoto, K. Tanaka, A. Fujimori, Y. Tomioka, and Y. Tokura, Phys. Rev. B {\bf 76}, 174418 (2007).
\bibitem{renner-cheong} Ch. Renner, et. al., Nature {\bf 416}, 518 (2002).
\bibitem{Biswas} Amlan Biswas, A K Raychaudhuri, R Mahendiran, A Guha, R Mahesh and C N R Rao, J. Phys.: Condens. Matter {\bf 9}, L355 (1997).
\bibitem{Kar} Sohini Kar, A K Raychaudhuri, APL {\bf 91}, 143124 (2007).
\bibitem{renner-bilayer}H. M. R\o nnow, Ch. Renner, G. Aeppli, T. Kimura, and Y. Tokura, Nature {\bf 440}, 1025 (2006).
\bibitem{Udai} U. R. Singh, S. Chaudhuri, S. K. Choudhary, R. C. Budhani and Anjan K. Gupta  Phys. Rev. B {\bf 77}, 014404 (2008).
\bibitem{Valencia} S. Valencia, A. Gaupp, W. Gudat, Ll. Abad, Ll. Balcells, and B. Martínez Phys. Rev. B {\bf 75}, 184431 (2007).
\bibitem{udai-apl} Udai Raj Singh, A. K. Gupta, G. Sheeth, V. Chandrashekhar, H. W. Jang, and C.-B. Eom, Appl. Phys. Lett. {\bf 93}, 212503 (2008).
\bibitem{res-note} We have also tried fitting to an activated form given by small polarons, i.e., $\rho(T)=\rho(T_0)Texp\left\{\frac{\Delta(T)}{k{_B}T}\right\}$. This form gives qualitatively the same variation of gap with T except for a smaller gap magnitude.
\bibitem{normalize-ref} J. A. Stroscio, R. M. Feenstra, and A. P. Fein, Phys. Rev. Lett. {\bf 57}, 2579 (1986).
\bibitem{Liu} H. L. Liu, S. L. Cooper, and S.-W. Cheong, Phys. Rev. Lett. {\bf 81}, 4684 (1998).
\bibitem{Park-Freeland} J.-H. Park, E. Vescovo, H.-J. Kim, C. Kwon, R. Ramesh, and T. Venkatesan, Phys. Rev. Lett. {\bf 81}, 1953 (1998); J. W. Freeland, K. E. Gray, L. Ozyuzer, P. Berghuis, E. Badica, J. Kavich, H. Zheng, and J. F. Mitchell, Nature Materials {\bf 4}, 62 (2005).
\bibitem{udai-jpcm} Udai Raj Singh, Saumyadip Chaudhary, R. C. Budhani, and A. K. Gupta, J. Phys.: Cond. Mat. {\bf 21}, 355001 (2009).
\bibitem{coh-pol} N. Mannella, W. L. Yang, X. J. Zhou, H. Zheng, J. F. Mitchell, J. Zaanen, T. P. Devereaux, N. Nagaosa, Z. Hussain, and Z.-X. Shen, Nature (London) 438, 474 (2005); Z. Sun, J. F. Douglas, A. V. Fedorov, Y. D. Chuang, H. Zheng, J. F. Mitchell, and D. S. Dessau, Nat. Phys. {\bf 3}, 248 (2007).
\bibitem{dagotto}R. Yu, S. Dong, C. Sen, G. Alvarez, and E. Dagotto, Phys. Rev. B {\bf 77}, 214434 (2008).
\end{thebibliography}
\end{document}